\documentclass[12pt,onecolumn]{IEEEtran}
\usepackage{amsmath,amssymb,amsthm,euscript,epsfig,times}

\newcommand{\pr}{{\mathbb{P}}}
\newcommand{\ex}{{\mathbb{E}}}

\newcommand{\openone}{\leavevmode\hbox{\small1\normalsize\kern-.33em1}}

\newtheorem*{thm}{Theorem}

\newtheorem*{defs}{Definition}

\DeclareMathOperator{\poly}{poly}

\title{Optimal Sequential Frame Synchronization
\thanks{This work was supported in part by NSF under Grant
No.~CCF-0515122, and by a University IR\&D Grant from Draper
Laboratory. }}
\author{Venkat Chandar, Aslan Tchamkerten, and Gregory Wornell\\
Electrical Engineering and Computer
  Science Department \\ Massachusetts Institute of Technology\\
Cambridge, MA 02139, USA \\
Email: $\{$vchandar,tcham,gww$\}$@mit.edu}

\begin{document}



%

\maketitle

\begin{abstract}  We consider the `one-shot frame synchronization problem'
where a decoder wants to locate a sync pattern at the output of a channel on
the basis of sequential observations. We assume that the sync pattern of length
$N$ starts being emitted at a random time within some interval of size $A$,
that characterizes the asynchronism level between the transmitter and the
receiver. We show that a sequential decoder can optimally locate the
sync pattern, i.e., exactly, without delay, and with probability approaching
one as $N\rightarrow \infty$, if and only if the asynchronism level grows as
$O(e^{N\alpha})$, with $\alpha$ below the {\emph{ synchronization threshold}},
a constant that admits a simple expression depending on the channel.  This
constant is the same as the one that characterizes the limit for reliable
asynchronous communication, as was recently reported by the authors. If $\alpha$
exceeds the synchronization threshold, any decoder, sequential or
non-sequential, locates the sync pattern with an error that tends to one as
$N\rightarrow \infty$. Hence, a sequential decoder can locate a sync pattern as well as the
(non-sequential) maximum likelihood decoder that operates on the basis of output sequences of maximum
length $A+N-1$, but with much fewer observations.

 \end{abstract}

{\keywords
Quickest detection, frame synchronization, sequential analysis}
\normalsize

\section{Introduction}
\label{intro}
Frame synchronization refers to the problem of locating a sync pattern
periodically embedded into data and received over a channel (see, e.g.,
\cite{Ma2,LT,Sc,Ni}). In \cite{Ma2}
Massey considered the situation of binary data transmitted across a white Gaussian noise
channel. He showed that, given received data of fixed size which the sync
pattern is known to belong to, the maximum likelihood rule consists of selecting the
location that maximizes the sum of the correlation and a correction term.

We are interested in the situation where the receiver wants to locate the sync
pattern on the basis of sequential observations, which Massey refers to as the
`one-shot' frame synchronization problem in \cite{Ma2}.  Surprisingly, this
setting has received much less attention than the fixed length frame setting. In
particular it seems that this problem hasn't been given a precise
formulation yet.
In this note we propose a formulation where the decoder has to
locate the sync pattern exactly and without delay, with the foreknowledge that
the pattern is sent within a certain time interval that characterizes the level
of asynchronism between the transmitter and the receiver. Our main result is
the asymptotic characterization of the largest asynchronism level with respect
to the size of the sync pattern for which a decoder can correctly perform with
arbitrarily high probability. 

\section{Problem formulation and result}
\label{pform}

We consider discrete-time communication over a discrete memoryless channel
characterized by its finite input
and output alphabets $\cal{X}$ and $\cal{Y}$, respectively, transition
probability matrix $Q(y|x)$, for all $y\in {\cal{Y}}$ and $x\in {\cal{X}}$, and
`noise' symbol $\star\in {\cal{X}}$.\footnote{Throughout this note we always assume that for all
$y\in {\cal{Y}}$ there is some $x\in {\cal{X}}$ for which $Q(y|x)>0$.}

The sync pattern $s^N$ consists of $N\geq 1$ symbols from $\cal{X}$ --- possibly also
the $\star$ symbol. The transmission of the sync pattern starts at a random time
$\nu$, uniformly distributed in $[1,2,\ldots,A]$, where the integer $A\geq 1$
characterizes the asynchronism level between the transmitter and the receiver.

We assume that the receiver knows $A$ but not $\nu$. Before and after the
transmission of the sync pattern, i.e., before time $\nu$ and after time
$\nu+N-1$, the receiver observes noise. Specifically, conditioned on the value
of $\nu$, the receiver observes independent symbols $Y_1,Y_2,\ldots$
distributed as follows. If $i\leq \nu-1$ or $i\geq \nu+N$, the distribution
is   $Q(\cdot|\star)$. At any time $i\in [\nu ,\nu+1,\ldots,  \nu+N-1]$ the
distribution is $Q(\cdot|{s_{i-\nu+1}})$, where $s_n$ denotes the $n$\/th
symbol of $s^N$.

\begin{figure} \begin{center} \input{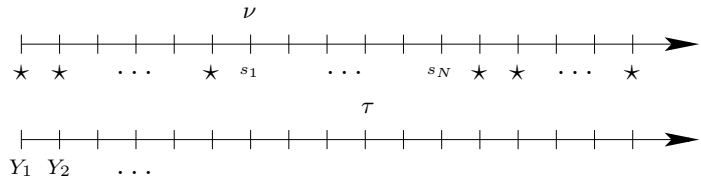}
\caption{\label{grapheesss} Time representation of what is sent (upper arrow)
and what is received (lower arrow). The `$\star$' represents the `noise'
symbol. At time $\nu$ the sync pattern starts being sent and is detected at time
$\tau$.} \end{center} \end{figure}

To identify the instant when the sync pattern starts being emitted, the receiver uses a
sequential decoder in the form of a stopping time $\tau$ with respect to the output sequence
$Y_1,Y_2,\ldots$\footnote{Recall that a stopping time $\tau$ is an
integer-valued random variable with respect to a sequence of random variables
$\{Y_i\}_{i=1}^\infty$ so that the event $\{\tau=n\}$, conditioned on
$\{Y_i\}_{i=1}^{n}$,  is independent of $\{Y_{i}\}_{i=n+1}^{\infty}$ for all
$n\geq 1$.} If $\tau=n$
the receiver declares that the sync pattern started being sent at time
$n-N+1$ (see
Fig.~\ref{grapheesss}).
The associated error probability is defined as
$$\pr(\tau\ne \nu+N-1)\;.$$

\noindent We now define the {\emph{synchronization threshold}}.
\begin{defs}\label{tasso} An {\emph{asynchronism exponent $\alpha$ is
achievable}} if there exists a sequence of pairs sync pattern/decoder
$\{(s^N,\tau_N)\}_{N\geq 1}$ such that $s^N$ and $\tau_N$ operate under
asynchronism level $A=e^{\alpha N}$, and so that $$\pr(\tau_N\ne
\nu-N+1)\overset{N\to \infty}{\longrightarrow}0\;.$$ The {\emph{synchronization
threshold}}, denoted $\alpha(Q)$, is the supremum of the set of achievable
asynchronism exponents.
\end{defs}
\noindent Our main result lies in the
following theorem.
\begin{thm}\label{unow}
 The synchronization threshold as
defined above is given by $$\alpha(Q)=\max_x D(Q(\cdot|x)||Q(\cdot|\star))\;,$$
where $D(Q(\cdot|x)||Q(\cdot|\star))$ is the divergence (Kullback-Leibler
distance) between $Q(\cdot|x)$ and $Q(\cdot|\star)$. Furthermore, if the
asynchronism exponent is above the synchronization threshold, a maximum
likelihood decoder that is revealed  the maximum length sequence of size
$A+N-1$ makes an error with a probability that tends to one as $N\rightarrow
\infty$.
\end{thm}
 A direct consequence of the theorem is that a sequential
decoder can (asymptotically) locate the sync pattern  as well as the optimal maximum likelihood
decoder that operates on a non-sequential basis, having access to sequences of
maximum size $A+N-1$.\footnote{Note that
if the receiver has no foreknowledge on $A$, i.e., if $A$ can a priori be
arbitrarily large, the problem is ill-posed: for any decoder, the probability of
miss-location of the sync pattern can be made arbitrarily large for $A$ large
enough.}


Note that the synchronization threshold is the same as the one in \cite{TCW},
which is defined as the largest asynchronism level for which reliable
communication can be achieved over point-to-point asynchronous channels.
This should not come as a surprise since the limit of asynchronous
communication is obtained in the zero rate regime where decoding errors are
mainly due to a miss location of the transmitted message.

We now prove the theorem by first presenting the direct part and then its
converse. 
Recall that a type, or empirical distribution, induced by a
sequence $z^N\in {\cal{Z}}^N$ is the probability $\hat{P}$ on $\cal{Z}$ where
$\hat{P}(a)$, $a\in {\cal{Z}}$, is equal to the number of occurrences of $a$ in $z^N$ divided by $N$. \begin{proof}[Proof of achievability] We
show that a suitable sync pattern together with the sequential typicality
decoder\footnote{\label{piednote}The sequential typicality decoder operates as
follows. At time $n$, it computes the empirical distribution
$\hat{P}$ induced by the sync pattern and the previous $N$
output symbols $y_{n-N+1},y_{n-N+2},\ldots,y_n$. If this distribution is close enough to 
$P$, i.e., if $|\hat{P}(x,y)-P(x,y)|\leq \mu$ for
all $x,y$, the decoder stops, and declares $n-N+1$ as the time the sync pattern
started being emitted. Otherwise it moves one step ahead and repeats the
procedure. Throughout the
argument we assume that $\mu$ is a negligible strictly positive quantity.}
achieves an asynchronism exponent arbitrarily close to the synchronization
threshold. The intuition is as follows. Let $\bar{x}$ be a `maximally divergent
symbol,' i.e., so that  $$D(Q(\cdot|\bar{x})||Q(\cdot|\star)) = \alpha(Q)\;.$$ 
Suppose the sync pattern consists of $N$ repetitions of $\bar{x}$. If we use
the sequential typicality decoding we already have almost all the properties we need. Indeed,
if $\alpha<\alpha(Q)$, with negligible probability the noise  generates a block
of $N$ output symbols that is jointly typical with the sync pattern. Similarly,
the block of output symbols generated by the sync pattern is jointly typical
with the sync pattern with high probability. The only problem occurs when a
block of $N$ output symbols is generated partly by noise and partly by the sync
pattern. Indeed, consider for instance the block of $N$ output symbols from
time $\nu-1$ up to $\nu+N-2$. These symbols are all generated according to the
sync pattern, except for the first. Hence, whenever the decoder observes this
portion of symbols, it makes an error with constant probability. The argument
extends to any fixed length shift.

The reason that the decoder is unable to locate the sync pattern exactly is
that a constant sync pattern has the undesirable property that when it is shifted to
the right, it still looks almost the same. Therefore, to prove the direct part
of the theorem, we consider a sync pattern mainly composed of
$\bar{x}$'s, but with a few $\star$'s mixed in\footnote{Indeed, any symbol different
than $\bar{x}$ can be used.} to make sure that shifts of the sync pattern
look sufficiently different from the original sync pattern. This allows the decoder to identify the sync
pattern exactly, with no delay, and with probability tending to one as $N$ goes
to infinity, for any asynchronism exponent less than
$\alpha(Q)$. We formalize this below.

Suppose that, for any arbitrarily large $K$, we can construct a sequence of patterns
$\{s^N\}$ of increasing lengths such that each $s^N=s_1,s_2,\ldots,s_N$ satisfies the
following two properties:
\begin{itemize}
\item[I.]
all $s_i$'s are equal to $\bar{x}$, except for a fraction smaller than $1/K$ that are
equal to $\star$;
\item[II.]
the Hamming distance between the pattern and any of its shifts of the form
$$\underbrace{\star, \star,\ldots, \star}_{i\text{ times}}, s_1,s_2,\ldots,s_{N-i}\;\;\quad i\in [1,2,\ldots,N]$$
 is linear in $N$.
\end{itemize}
Now let $A = e^{N (\alpha(Q) - \epsilon)}$,
for some $\epsilon > 0$, and consider using patterns with the properties I
and II in conjunction with the sequential typicality
decoder $\tau=\tau_N$.

By \cite[Lemma 2.6, p.32]{CK} and property I, the probability that $N$ output symbols entirely generated by noise are typical
with the sync pattern is upper bounded by
$\exp \left(- N
(1-1/K)(\alpha(Q)-\delta)\right)$,
 where $\delta>0$ goes to zero
as the typicality constant $\mu$ goes to zero.\footnote{See footnote \ref{piednote}.} Hence, by the
union bound
$$\pr\left(\{\tau<\nu\right\}\cup \left\{\tau\geq \nu+2N-1\}\right) \leq e^{- N
(\epsilon-\delta-(\alpha-\delta)/K)}$$
which tends to zero for $\mu$ small enough and $K$ sufficiently large.\footnote{If $\alpha(Q)=\infty$ the
upper bound is zero if $\mu$ is small enough.}

If the $N$ observed symbols are partly generated by noise and partly by the
sync pattern, by property II, the Chernoff bound, and the union bound we obtain
$$\pr\left( \tau \in [\nu,\nu+1,\ldots,
\nu+N-2]\right) \leq (N-1)e^{-
\Omega(N)}$$  which vanishes as $N$ tends to infinity.

We then deduce that  $$\pr(\tau=\nu+N-1) \rightarrow
1$$ as $N \rightarrow \infty$.

To conclude we give an explicit construction of a sequence of sync pattern
satisfying the properties I and II above. To that aim we use maximal
length shift register sequences (see, e.g.,\cite{Gol}). Actually, for our
purpose, the only property we use from such binary sequences of length
$l=2^m-1$, $m\in [1,2,\ldots)$, is that they are of Hamming distance $(l+1)/2$
from any of their circular shifts.

To construct the sync pattern we start by setting $s_i = \bar{x}$ for all $i
\not\equiv 0$ mod $K$ where, without loss of generality, $K$ is chosen to
satisfy $\lfloor \frac{N}{K} \rfloor=2^m-1$
for some $m\in [1,2,\ldots)$.\footnote{We use $\left\lfloor x\right\rfloor$
to denote the largest integer smaller than $x$.} With this choice, property I
is already satisfied. To specify the $\lfloor \frac{N}{K} \rfloor$ positions
$i$ with $i \equiv 0$ mod $K$, pick a maximal length shift register sequence
$m_1,m_2,\ldots, m_{\lfloor \frac{N}{K} \rfloor}$, and set
 $s_{jK} = \bar{x}$ if $m_j = 0$ and $s_{jK} = \star$ if $m_j = 1$, for any
integer $j\leq \lfloor \frac{N}{K} \rfloor$. It can be readily verified, using
the circular shift property of maximal length shift register sequences, that
this construction yields patterns that satisfy property~II.

 \end{proof}

\begin{proof}[Proof of the converse]
We assume that $A=e^{N\alpha}$
with $$\alpha>\max_x
D(Q(\cdot|x)||Q(\cdot|\star))$$
and show that the (optimal) maximum likelihood decoder that operates on the
basis of sequences of maximum length $A+N-1$
yields a probability of error going to one as $N$ tends to infinity.

We assume that the sync pattern $s^N$ is composed of $N$ identical symbols
$s\in \cal{X}$. The case with multiple symbols is obtained by a straightforward
extension. Suppose the maximum likelihood decoder not only is revealed the complete
sequence $$y_1,y_2,\ldots,y_{A+N-1}\;,$$ but also knows that the sync pattern was sent
in one of the $r$ distinct block periods of duration $N$, where $r$ denotes the
integer part of $(A+N-1)/N$, as shown in Fig.~\ref{graphees2}.

\begin{figure}
\begin{center}
\input{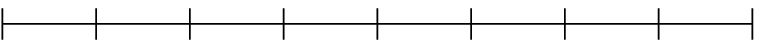}
\caption{\label{graphees2} Parsing of the entire received sequence of size $A+N-1$
into $r$ blocks $y^{(t_1)},y^{(t_2)},\ldots,y^{(t_r)}$ of
length $N$, where the $i$\/th block starts at time $t_i$.}
\end{center}
\end{figure}

Assuming $Q(y|\star)>0$ for all $y\in \cal Y $,\footnote{If $Q(y|\star)=0$ for
some $y\in \cal Y $ we have $\alpha(Q)=\infty$, and there is nothing to prove.}
straightforward algebra shows that the decoder outputs the time $t_i$,
$i\in [1,2,\ldots,r]$, that maximizes $$f(y^{(t_i)})=
\frac{Q(y^{(t_i)}|s^N)}{Q(y^{(t_i)}|\star)}\;.$$ Note that $f(y^{(t)})$ depends
only on the type of the sequence $y^{(t)}$ since $s^N$ is the repetition of a
single symbol. For conciseness, from now on we adopt the notation
$Q_s(y^{(t)})$ instead of $Q(y^{(t)}|s^N)$ and $Q_\star(y^{(t)})$ instead of
$Q(y^{(t)}|\star)$.

Let $Q_s\pm \varepsilon_0$ denote the set of types (induced by sequences $y^N$) that are
$\varepsilon_0>0$ close to $Q_s$ with respect to the $L_1$ norm, and let $E_1$
denote the event that the type of the $\nu$\/th block  (corresponding to the
sync transmission period) is not
in $Q_s\pm \varepsilon_0$. It
follows that
\begin{align}\label{e1}
\pr(E_1) \leq
e^{-N\epsilon}
\end{align}
 for some $\varepsilon=\varepsilon(\varepsilon_0)>0$.\footnote{Here we
implicitly assume that $N$ is large enough so that the set of types $Q_s\pm
\varepsilon_0$ is nonempty. }
Let
$\bar{Q_s} = \arg \max_{P\in Q_s\pm \epsilon_0} f(P)$,\footnote{Note that
$\bar{Q_s}$ may not be equal to $Q_s$.} where with a slight abuse of notation
$f(P)$ is used to denote $f(y^N)$ for any sequence $y^N$ having type $P$. Now
consider the event $E_2$ where the number of blocks generated by $Q_\star$ that have
type $\bar{Q}_s$ is
smaller than
$$\frac{1}{2(N+1)^{|\cal{X}|}}e^{-N(D(\bar{Q_s}||Q_\star)-\alpha)}\;.$$
Using \cite[Lemma 2.6, p.32]{CK}, the expected number of blocks generated by
$Q_\star$ that have type
$\bar{Q_s}$ is lower bounded as

\begin{align*} \ex\left(\mbox{number of type $\bar{Q_s}$ blocks generated from $Q_\star$}\right)&\geq \frac{1}{(N+1)^{|\cal{X}|}}e^{-N
D(\bar{Q_s}||Q_\star)}(r-1)\\ &\geq \poly(N)e^{-N
(D(\bar{Q_s}||Q_\star)-\alpha)}\;,
\end{align*}
and using Chebyshev's inequality we get
\begin{align}\label{e2} \pr\big(E_2\big)\leq \text{poly}(N){e^{-N(\alpha-D(\bar{Q_s}||Q_\star))}}
\end{align}
where poly$(N)$ denotes a term that increases or decreases no faster than
polynomially in $N$.

Finally consider the event $E_3$ defined as the complement of
$E_1\cup E_2$. Given that $E_3$ happens, the decoder
sees at least
$$\frac{1}{2(N+1)^{|\cal{X}|}}e^{-N(D(\bar{Q_s}||Q_\star)-\alpha)}$$
time slots that are at least as probable as the correct $\nu$\/th.
Hence, the probability of correct detection given that the event $E_3$ happens
is upper bounded as
\begin{align}\label{e3}
\pr(\text{corr.dec}|E_3)\leq \text{poly}(N)e^{-N(\alpha -D(\bar{Q_s}||Q_\star))}\;.
\end{align}
We deduce from \eqref{e1}, \eqref{e2}, and \eqref{e3}
that the probability of correct decoding is upper bounded as
\begin{align*} \pr\left(\text{corr. dec.}\right)&=\sum_{i=1}^3\pr(\text{corr.dec}|E_i)\pr(E_i)\nonumber\\
&\leq \pr(E_1)+\pr(E_2)+\pr(\text{corr.dec}|E_3)\\
&\leq (e^{-N\varepsilon}+{e^{-N(\alpha -D(\bar{Q_s}||Q_\star))}})\text{poly}(N)\;. \end{align*}
Therefore if
 $$\alpha> D(\bar{Q_s}||Q_\star)\;,$$
the probability of successful detection goes to zero as $N$ tends to infinity.
Since $D(\bar{Q_s}||Q_\star)$ tends to $D({Q_s}||Q_\star)$ as
$\varepsilon_0\downarrow 0$
by continuity of $D(\cdot||Q_\star)$,\footnote{We may assume $D(\cdot||Q_\star)$ is continuous because otherwise $\alpha(Q) = \infty$ and there
is nothing to prove.}
the result follows by maximizing $D(Q_s||Q(\cdot|\star))$ over $s\in {\cal{X}}$.

\end{proof}

\bibliographystyle{amsplain}
\bibliography{../../common_files/bibiog}

\end{document}